\documentclass[12pt]{article}
\usepackage{graphicx}
\usepackage[cp1251]{inputenc}
\usepackage{epsfig}
\usepackage[english]{babel}

\date{}

\begin{document}
\setcounter{page}{1}
\pagestyle{plain}

\title{\bf{Strain engineering the charged-impurity-limited carrier mobility in phosphorene}}

\author{Yawar Mohammadi$^1$\thanks{Corresponding author. Tel./fax: +98 831 427
4569, Tel: +98 831 427 4569. E-mail address:
yawar.mohammadi@gmail.com} , Borhan Arghavani Nia$^{2}$}
\maketitle{\centerline{$^1$Young Researchers and Elite Club,
Kermanshah Branch, Islamic Azad University, Kermanshah, Iran}
\maketitle{\centerline{$^2$Department of Physics, Kermanshah
Branch, Islamic Azad University, Kermanshah, Iran}

\begin{abstract}

We investigate, based on the tight-binding model and in the linear
deformation regime, the strain dependence of the electronic band
structure of phosphorene, exposed to a uniaxial strain in one of
its principle directions, the normal, the armchair and the zigzag
directions. We show that the electronic band structure of strained
phosphorene, for the experimentally accessible carrier densities
and the uniaxial strains, is well described by a strain-dependent
decoupled electron-hole Hamiltonian. Then, employing the decoupled
Hamiltonian, we consider the strain dependence of the
charged-impurity-limited carrier mobility in phosphorene, for both
types of carriers, arbitrary carrier densities and in both
armchair and zigzag directions. We show that a uniaxial tensile
(compressive) strain in the normal direction enhances (weakens)
the anisotropy of the carrier mobility, while a uniaxial strain in
the zigzag direction acts inversely. Moreover applying a uniaxial
strain in the armchair direction is shown to be ineffective on the
anisotropy of the carrier mobility. These will be explained based
on the effects of the strains on the carrier effective masses.
\end{abstract}

%{\it \emph{PACS}}: \emph{74.20.-z, 74.20.Fg}

\vspace{0.5cm}

{\it \emph{Keywords}}: Phosphorene; Tight-binding model; Band
Structure; Strain; Carrier Mobility.
%
%\newpage
\section{Introduction}
\label{sec:1}

Since successful isolation of a single layer of
graphite\cite{Novoselov1} called graphene, as the first real
two-dimensional lattice structure which shows novel appealing
properties\cite{Neto1,Peres1}, many researchers tried to synthesis
or isolate new two-dimensional materials. These efforts resulted
in finding other two dimensional materials such as
BN\cite{Pacile1}, transition metal
dichalcogenides\cite{Novoselov2},
silicene\cite{Lalmi1,Padova1,Aufray1,Vogt1} and recently
phosphorene. Phosphorene is a single layer of black phosphorus,
which can be isolated by mechanical exfoliation\cite{Liu1,Li1} of
black phosphorus. In a single layer of black phosphorus, each
phosphorus atom covalently couples to three nearest neighbors.
This configuration of phosphorus atoms results in a honeycomb-like
lattice structure. However, due to the $sp^{3}$ hybridization of
$s$ and $p$ atomic orbitals, it forms a puckered surface. The
electronic band structure of phophorene has been studied using
different methods such density functional theory
calculations\cite{Rodin1,Rudenko1}, $\mathbf{k}.\mathbf{p}$
method\cite{Xia1,Pengke1} and tight-binding
model\cite{Rudenko1,Ezawa1,Pereira1}. These considerations show
phophorene is a direct-band-gap insulator, but with an anisotopic
band structure. This novel band structure leads to many attractive
properties\cite{Qin1,Low1,Ong1,Jiang1}.

Strain tuning is an effective means to tune the physical
properties of two dimensional materials (for a review, see e.g.
Ref.\cite{Neto1,Amorim1}). Puckered structure of phosphorene makes
this easier, so one can tune and control its electronic and
mechanical properties by strain, confirmed by recent studies on
the effects of uniaxial and biaxial strains in phosphorene
\cite{Rodin1,Wei1,Fei1,Sa1,Jiang2,Han1,Peng1,Elahi1,Fei2,Cakir1,Ju1,Wang1}.
These works examined effects of strains applied along three
principle directions which preserve $D_{2h}$ group point symmetry
of phosphorene\cite{Ezawa1}, its zigzag and armchair edges and the
direction normal to its plane. It has been shown that a uniaxial
strain in the direction normal can decrease its band gap and even
leads to an insulator to metal transition\cite{Rodin1,Han1,Qin2}.
Moreover, the effects of an in-plane uniaxial strains along zigzag
and armchair edges\cite{Peng1,Elahi1,Fei2} on the band gap of
phosphorene has been studied. Some other researchers has studied
the effects of uniaxial and biaxial strains on the band
structure\cite{Fei2,Cakir1,Ju1,Wang1} and the optical
properties\cite{Cakir1} of phosphorene, confirming the capability
of stain as an effective means to tune the properties of
phosphorene. These works showed, when the uniaxial strain is
applied along the armchair direction, the properties of phophorene
change further. But, recently, it has been shown\cite{Jiang3} that
the most effective direction to apply a strain and tune the band
gap of phosphorene is an in-plane direction (not being along
armchair nor along zigzag) with a direction angle about $0.268\pi$
counted from the armchair edge.

According to the high-potential capability of strains to tune the
properties of phosphorene, driving an analytical relation for the
Hamiltonian of strained phosphorene is very desirable, and can be
used to examine the effects of the strains on the electric,
optical and magnetic properties of phosphorene. In this paper,
starting from the well-known 4-band tight-binging Hamiltonian of
phophorene\cite{Rudenko1,Ezawa1}, we obtain a strain-dependent
tight-binding Hamiltonian for phosphorene. In this paper we work
in the linear deformation regime and only consider uniaxial
strains. To benefit from the $D_{2h}$ group point symmetry of
phosphorene and reduce the 4-band Hamiltonian to a 2-band
Hamiltonian and achieve an analytical result, we restrict our
consideration to the uniaxial strains applied along three
principle directions of phosphorene which preserve its $D_{2h}$
group point symmetry. Thanks to this symmetry, we can obtain
analytical relations for its band energies which can be used to
explore easily the effects of the uniaxial strains on properties
of phosphorene. Searching for low-energy structures in strained
phosphorene, we use continuum approximation and derive the
corresponding Hamiltonian dominating low energy excitations. Then,
by taking into account the weak interband coupling of conduction
and valance bands, we project the low-energy Hamiltonian into a
decoupled Hamiltonian\cite{Zhou1,Gillgren1,Tahir1} and show that
for the experimentally accessible carrier density, the decoupled
bands agree well the bands obtained from the tight-binding
Hamiltonian. Motivated by this fact and the recent studies on the
carrier mobility in phopsphorene\cite{Liu1,Ong1,Gillgren1,Fei2},
then we apply our decoupled Hamiltonian to consider the strain
dependence of the charged-impurity-limited carrier mobility in
phosphorene. Our result shows that one can tune the amount and the
anisotropy of the mobility in phosphorene, by making use of a
uniaxial strain in the normal or zigzag direction.

The rest of this paper is organized as follows. In Sec. II we
reproduce the known 4-band Hamiltonian of phosphorene. In Sec. III
we explain how one can, in general, insert the effects of the
strains in the Hamiltonian and obtain a general formalism for the
strain-dependent Hamiltonian of phosphorene. Sec. IV devoted to
consider the strain dependence of the charged-impurity-limited
carrier mobility in phosphorene. We end the paper by summarizing
our results in Sec. V.

\section{Structure and tight-binding Hamiltonian of phosphorene}
\label{sec:2}

The lattice structure of phosphorene and the necessary lattice
parameters to construct the tight-binding Hamiltonian of
phosphorene, including the lattice constant, the bond angles and
the transfer energies, have been introduced in Fig. \ref{fig.01}.
The unit cell of phosphorene (solid-line rectangle in Fig.
\ref{fig.01}) consists of four phophorus atoms, two atoms in the
lower layer represented by the grey circles (called $A$ and $B$)
and two atoms in upper layer represented by the red circles
(called $C$ and $D$). Hence, the tight-binding Hamiltonian of
phophorene can be written in terms of a $4\times4$ matrix as
\begin{eqnarray}
\widehat{H}_{\mathbf{k}} = \left(
\begin{array}{cccccc}
     0              & t_{AB}(\mathbf{k}) & t_{AC}(\mathbf{k}) & t_{AD}(\mathbf{k}) \\
 t_{BA}(\mathbf{k}) &         0          & t_{BC}(\mathbf{k}) & t_{BD}(\mathbf{k}) \\
 t_{CA}(\mathbf{k}) & t_{CB}(\mathbf{k}) &         0          & t_{CD}(\mathbf{k}) \\
 t_{DA}(\mathbf{k}) & t_{DB}(\mathbf{k}) & t_{DC}(\mathbf{k}) &         0          \\
\end{array}
\right),\label{eq.1}
\end{eqnarray}
acting in ${(\phi_{A},\psi_{B},\psi_{C},\psi_{D})}^{T}$ with
$\mathbf{k}$ being the two-dimensional momentum. Notice that
$t_{BA}(\mathbf{k})=t^{\ast}_{AB}(\mathbf{k})$. Moreover, it has
been shown\cite{Rudenko1,Ezawa1} that if we only retain the
transfer energies up to the fifth nearest neighbors, the
tight-binding approximated band structure agrees well with its
density functional theory band structure. These transfer energies
are\cite{Rudenko1} $t_{1}=-1.220~eV$, $t_{2}=+3.665~eV$,
$t_{3}=-0.205~eV$, $t_{4}=-0.105~eV$ and $t_{5}=-0.055~eV$. So we
can rewrite the Hamiltonian matrix as
\begin{eqnarray}
\widehat{H}_{\mathbf{k}} = \left(
\begin{array}{cccc}
     0             & f_{1\mathbf{k}}+f_{3\mathbf{k}} & f_{4\mathbf{k}} & f_{2\mathbf{k}}+f_{5\mathbf{k}} \\
f^{\ast}_{1\mathbf{k}}+f^{\ast}_{3\mathbf{k}} &      0      & f_{2\mathbf{k}}+f_{5\mathbf{k}} & f_{4\mathbf{k}} \\
f^{\ast}_{4\mathbf{k}} & f^{\ast}_{2\mathbf{k}}+f^{\ast}_{5\mathbf{k}} & 0 & f_{1\mathbf{k}}+f_{3\mathbf{k}} \\
f^{\ast}_{2\mathbf{k}}+f^{\ast}_{5\mathbf{k}} & f^{\ast}_{4\mathbf{k}} & f^{\ast}_{1\mathbf{k}}+f^{\ast}_{3\mathbf{k}} &  0 \\
\end{array}
\right),\label{eq.2}
\end{eqnarray}
where the matrix elements are given by
$f_{1\mathbf{k}}=2t_{1}e^{ik_{x}x_{1}}\cos(k_{y}y_{1})$,
$f_{2\mathbf{k}}=t_{2}e^{ik_{x}x_{2}}$,
$f_{3\mathbf{k}}=2t_{3}e^{ik_{x}x_{3}}\cos(k_{y}y_{3})$,
$f_{4\mathbf{k}}=4t_{4}\cos(k_{x}x_{4})\cos(k_{y}y_{4})$ and
$f_{5\mathbf{k}}=t_{5}e^{ik_{x}x_{5}}$. Here
$\vec{r}_{i}=(x_{i},y_{i},z_{i})$ is a vector which is drawn from
$A$ (The origin of the cartesian coordinate system) to one of the
ith nearest neighbors (See Fig. \ref{fig.01}). They are
$\vec{r}_{1}=(-d_{1}\cos\alpha,d_{1}\sin\alpha,0)$,
$\vec{r}_{2}=(d_{2}\cos\theta,0,d_{2}\sin\theta)$,
$\vec{r}_{3}=(d_{1}\cos\alpha+2d_{2}\cos\theta,d_{1}\sin\alpha,0)$,
$\vec{r}_{4}=(-d_{1}\cos\alpha-d_{2}\cos\theta,d_{1}\sin\alpha,d_{2}\sin\theta)$
and
$\vec{r}_{5}=(-2d_{1}\cos\alpha-d_{2}\cos\theta,0,d_{2}\sin\theta)$
where $\cos\theta=-\frac{\cos\beta}{\cos\alpha}$. One can take
into account the $D_{2h}$ group point symmetry in phosphorene and
project the four-band Hamiltonian into a reduced two-band
Hamiltonian as\cite{Ezawa1}
\begin{eqnarray}
\widehat{H}_{\mathbf{k}} = \left(
\begin{array}{cc}
          f_{4\mathbf{k}}         &        f_{1\mathbf{k}}+f_{2\mathbf{k}}+f_{3\mathbf{k}}+f_{5\mathbf{k}}  \\
f^{\ast}_{1\mathbf{k}}+f^{\ast}_{2\mathbf{k}}+f^{\ast}_{3\mathbf{k}}+f^{\ast}_{5\mathbf{k}} &      f_{4\mathbf{k}}  \\
\end{array}
\right),\label{eq.3}
\end{eqnarray}
acting in $(\phi_{A}+\phi_{C},\phi_{B}+\phi_{D})^{T}/2$. The
corresponding energy bands, obtained by diagonalizing the
Hamiltonian matrix, are given by
\begin{eqnarray}
E_{\mathbf{k}}=f_{4\mathbf{k}}\pm|f_{1\mathbf{k}}+f_{2\mathbf{k}}+f_{3\mathbf{k}}+f_{5\mathbf{k}}|,\label{eq.4}
\end{eqnarray}
where $+(-)$ denotes to the conduction(valance) band. We have
shown the energy spectrum of phosphorene obtained from two-band
Hamiltonian in Fig. \ref{fig.02}. It is evident that minimum
(maximum) of the conduction (valance) energy band is at $\Gamma$
point. If we apply continuum approximation to the obtained
two-band Hamiltonian and retain the terms up to the second order
in k, we can reproduce the known Hamiltonian of
phosphorene\cite{Ezawa1,Pereira1},
\begin{eqnarray}
\widehat{H}_{\mathbf{k}} = \left(
\begin{array}{cc}
 u+\eta_{x}k^{2}_{x}+\eta_{y}k^{2}_{y}            & \delta+\gamma_{x}k^{2}_{x}+\gamma_{y}k^{2}_{y}+i\chi k_{x}  \\
 \delta+\gamma_{x}k^{2}_{x}+\gamma_{y}k^{2}_{y}-i\chi k_{x} &  u+\eta_{x}k^{2}_{x}+\eta_{y}k^{2}_{y}   \\
\end{array}
\right),\label{eq.5}
\end{eqnarray}
where $u=4t_{4}=0.42~eV$, $\eta_{x}=-2t_{4}x^{2}_{4}=1.03~eV
\AA^{2}$, $\eta_{y}=-2t_{4}y^{2}_{4}=0.56~eV \AA^{2}$,
$\delta=2t_{1}+t_{2}+2t_{3}+t_{5}=0.76~eV$,
$\gamma_{x}=-t_{1}x_{1}^{2}-\frac{t_{2}}{2}x_{2}^{2}-t_{3}x_{3}^{2}-\frac{t_{5}}{2}x_{5}^{2}=3.51~eV
\AA^{2}$, $\gamma_{y}=-t_{1}y_{1}^{2}-t_{3}y_{3}^{2}=3.81~eV
\AA^{2}$ and
$\chi=2t_{1}x_{1}+t_{2}x_{2}+2t_{3}x_{3}+t_{5}x_{5}=-5.34~eV \AA$
which agree well with the other calculations
\cite{Pereira1}(Notice that in our calculations the zigzag edge
lies along the x-axis.). The corresponding energy spectrums are
given by
\begin{eqnarray}
E_{\mathbf{k}}=u+\eta_{x}k^{2}_{x}+\eta_{y}k^{2}_{y}\pm\sqrt{(\delta+\gamma_{x}k^{2}_{x}+\gamma_{y}
k^{2}_{y})^{2}+\chi^{2}k_{x}^{2}},\label{eq.6}
\end{eqnarray}
where $+(-)$ denotes to the conduction(valance) band. It is
evident that the energy spectrum is linear in the $k_{x}$
direction while in the $k_{y}$ direction it is parabolic. Due to
the large band gap, which leads to a weak interbans coupling, one
can decouple the electron and the hole bands into a low energy
regime. In this approximation, Eq. \ref{eq.6} can be written
as\cite{Pereira1,Zhou1,Gillgren1,Tahir1}
\begin{eqnarray}
E_{\mathbf{k}}\approx u+\eta_{x}k^{2}_{x}+\eta_{y}k^{2}_{y} \pm
\delta(1+\frac{1}{2}[2\frac{\gamma_{x}}{\delta}k^{2}_{x}+2\frac{\gamma_{y}}{\delta}
k^{2}_{y}+\frac{\chi^{2}k_{x}^{2}}{\delta^{2}}]).\label{eq.7}
\end{eqnarray}
In this approximation the electron and hole effective masses in
the $x$ and $y$ directions are given by,
$m_{ex}=\frac{\hbar^{2}}{2(\eta_{x}+\gamma_{x}+\chi^{2}/2\delta)}=0.168~m_{0}$,
$m_{ey}=\frac{\hbar^{2}}{2(\eta_{y}+\gamma_{y})}=0.852~m_{0}$,
$m_{hx}=\frac{\hbar^{2}}{2(\gamma_{x}-\eta_{x}-\chi^{2}/2\delta)}=0.184~m_{0}$
and $m_{hy}=\frac{\hbar^{2}}{2(\gamma_{y}-\eta_{y})}=1.146~m_{0}$,
which $m_{0}$ is the mass of a free electron, in good agreement
with recent result\cite{Rudenko1}. To see that in what region this
approximated energy bands agree well with the other results, we
have shown all three set energy bands obtained from the
tight-binding, the low energy and the decoupled Hamiltonians, in
Fig. \ref{fig.02}. One can see that in the $k_{y}$ direction all
three set energy bands agree well in a wide range of the energy
and the momentum. Moreover, in the $k_{x}$ direction the
tight-binding energy bands and the low-energy bands agree well
too, but the decoupled bands overlap with them only up to
$0.14~eV$ ($0.13~eV$) with respect to the bottom (top) of the
conduction (valance) bands. This corresponds to
$n=2.20\times10^{13}~cm^{-2}$ and $n=2.44\times10^{13}~cm^{-2}$
for the electron and hole densities. These indicate that the low
energy excitations in phosphorene are well described by the
decoupled Hamiltonian\cite{Pereira1,Zhou1,Gillgren1,Tahir1}.

\section{Strain-dependent tight-binding Hamiltonian}
\label{sec:3}

In this section we rederive the tight-binding Hamiltonian of
phosphorene in the presence of the uniaxial strains applied along
the principle directions of phosphorene. To insert the effects of
the applied strain in the tight-binding Hamiltonian of
phosphorene, first we must determine the effects of the strain on
the transfer energies and the bond lengths. It has been
shown\cite{Harrison1} that the transfer energies between $s$ and
$p$ orbitals, which construct the electronic bands of phosphorene,
depend on the bond length as $t\propto\frac{1}{r^{2}}$. To obtain
this relation, it has been supposed that the applied strain
doesn't change the bond angles and only change the bond lengths.
Within the linear deformation regime, this is a reasonable
assumption. Since the change in the bond angles in a strained
lattice, at least, includes the terms of second order in terms of
the applied strain. So they can be ignored in the linear
deformation regime. Hence, we only need to determine the strain
dependence of the bond lengths and the other inter-atomic
distances.

Let us construct our formalism in a general case in which
phosphorene is exposed to strains applied along all three
principle directions of phosphorene, the armchair (x-direction)
and the zigzag (y-direction) edges and the the direction normal to
the phosphorene plane (z-direction). So, the deformed coordinates
are given by
\begin{eqnarray}
\left(
\begin{array}{c}
 x^{\epsilon}  \\
 y^{\epsilon}  \\
 z^{\epsilon}  \\
\end{array}
\right) = \left(
\begin{array}{ccc}
 1+\epsilon_{x}  &        0       &         0      \\
        0        & 1+\epsilon_{y} &         0      \\
        0        &        0       & 1+\epsilon_{z} \\
\end{array}
\right) \left(
\begin{array}{c}
 x \\
 y  \\
 z \\
\end{array}
\right),\label{eq.8}
\end{eqnarray}
where $\epsilon_{x}$, $\epsilon_{y}$ and $\epsilon_{z}$ are the
normal strains applied along the x-, y- and z-directions
respectively. In this paper, we restrict our considerations to the
linear deformation regime, so the bond lengths and the other
atomic distances, in general, can be expanded in terms of all
components, $\epsilon_{x}$, $\epsilon_{y}$ and $\epsilon_{z}$ as
\begin{eqnarray}
r^{\epsilon}=r+\alpha_{x}\epsilon_{x}+\alpha_{y}\epsilon_{y}+\alpha_{z}\epsilon_{z},\label{eq.9}
\end{eqnarray}
where $r$ and
$r^{\epsilon}=\sqrt{(x^{\epsilon})^{2}+(y^{\epsilon})^{2}+(z^{\epsilon})^{2}}$
are the unstrained and strained bond lengths respectively and
$\alpha_{x}$, $\alpha_{y}$ and $\alpha_{z}$ are the strain-related
geometrical coefficients, given by $\alpha_{x}=\frac{\partial
r^{\epsilon}}{\partial x}|_{\epsilon_{x}=0}=\frac{x^{2}}{r}$,
$\alpha_{y}=\frac{\partial r^{\epsilon}}{\partial
y}|_{\epsilon_{y}=0}=\frac{y^{2}}{r}$ and
$\alpha_{z}=\frac{\partial r^{\epsilon}}{\partial
z}|_{\epsilon_{z}=0}=\frac{z^{2}}{r}$. If we insert Eq. \ref{eq.9}
into the relation $t\propto\frac{1}{r^{2}}$, and expand it in
terms of the strains and only retain the terms up to the first
order in $\epsilon$ we get
\begin{eqnarray}
t^{\epsilon}=t-\frac{2}{r}(\alpha_{x}\epsilon_{x}+\alpha_{y}\epsilon_{y}+\alpha_{z}\epsilon_{z})t,\label{eq.10}
\end{eqnarray}
where $t$ and $t^{\epsilon}$ are the unstrained and the strained
transfer energies respectively.

As mentioned above, the strains applied along all three principle
directions of phosphorene don't break $D_{2h}$ symmetry. So the
electronic excitations in strained phosphorene are dominated by
the reduced two-band Hamiltonian, Eq. \ref{eq.3}, but after
substituting the deformed transfer energies and bond lengths into
it. Recalling the relations obtained for the strained bond lengths
and the transfer energies, and substituting them into the tow-band
Hamiltonian, Eq.~\ref{eq.3}, we get
\begin{eqnarray}
\widehat{H}_{\mathbf{k}} = \left(
\begin{array}{cc}
 f^{\epsilon}_{4\mathbf{k}}  &   f^{\epsilon}_{1\mathbf{k}}+f^{\epsilon}_{2\mathbf{k}}+
 f^{\epsilon}_{3\mathbf{k}}+f^{\epsilon}_{5\mathbf{k}}  \\
 f^{\epsilon\ast}_{1\mathbf{k}}+f^{\epsilon\ast}_{2\mathbf{k}}+f^{\epsilon\ast}_{3\mathbf{k}}+
 f^{\epsilon\ast}_{5\mathbf{k}} &  f^{\epsilon}_{4\mathbf{k}}  \\
\end{array}
\right),\label{eq.11}
\end{eqnarray}
where
$f_{1\mathbf{k}}=2t^{\epsilon}_{1}e^{ik_{x}x^{\epsilon}_{1}}\cos(k_{y}y^{\epsilon}_{1})$,
$f_{2\mathbf{k}}=t^{\epsilon}_{2}e^{ik_{x}x^{\epsilon}_{2}}$,
$f_{3\mathbf{k}}=2t^{\epsilon}_{3}e^{ik_{x}x_{3}}\cos(k_{y}y^{\epsilon}_{3})$,
$f_{4\mathbf{k}}=4t^{\epsilon}_{4}\cos(k_{x}x^{\epsilon}_{4})\cos(k_{y}y^{\epsilon}_{4})$
and $f_{5\mathbf{k}}=t^{\epsilon}_{5}e^{ik_{x}x^{\epsilon}_{5}}$.
The corresponding electron and hole energy bands are given by
\begin{eqnarray}
E^{\epsilon}_{\mathbf{k}}=f^{\epsilon}_{4\mathbf{k}}\pm|f^{\epsilon}_{1\mathbf{k}}+f^{\epsilon}_{2\mathbf{k}}+
f^{\epsilon}_{3\mathbf{k}}+f^{\epsilon}_{5\mathbf{k}}|.\label{eq.12}
\end{eqnarray}
with $+(-)$ denoting to the electron (hole) band. It easy to show
that strained phopsphorene has a direct band gap at $\Gamma$
point, in agreement with the recent density functional theory
\cite{Rodin1,Han1,Qin2,Peng1} and tight-binding\cite{Jiang4}
calculations done in the linear deformation regime. Similar to the
unstrained case, to capture the low energy physics of strained
phosphorene, one can expand the matrix elements around $\Gamma$
point and only retain the terms up to second order in $k$ and
first order in $\epsilon$. Hence, the low energy Hamiltonian
becomes
\begin{eqnarray}
\widehat{H}_{\mathbf{k}} = \left(
\begin{array}{cc}
 u^{\epsilon}+\eta^{\epsilon}_{x}k^{2}_{x}+\eta^{\epsilon}_{y}k^{2}_{y}   &
 \delta^{\epsilon}+\gamma^{\epsilon}_{x}k^{2}_{x}+\gamma^{\epsilon}_{y}k^{2}_{y}+i\chi^{\epsilon} k_{x}  \\
 \delta^{\epsilon}+\gamma^{\epsilon}_{x}k^{2}_{x}+\gamma^{\epsilon}_{y}k^{2}_{y}-i\chi^{\epsilon}k_{x}   &
 u^{\epsilon}+\eta^{\epsilon}_{x}k^{2}_{x}+\eta^{\epsilon}_{y}k^{2}_{y}   \\
\end{array}
\right),\label{eq.13}
\end{eqnarray}
The values of the matrix elements depends on the directions in
which the strains are applied (See appendix). The applied strains
can affect, by changing the energy gap and $\chi^{\epsilon}$, on
the interband coupling. When the interband coupling is weak, one
can project the low energy two-band Hamiltonian into a decoupled
Hamiltonian which is given by
\begin{eqnarray}
\widehat{H}_{\mathbf{k}} = \left(
\begin{array}{cc}
 E^{\epsilon}_{e}+\frac{\hbar^{2}k^{2}_{x}}{2m^{\epsilon}_{ex}}+\frac{\hbar^{2}k^{2}_{y}}{2m^{\epsilon}_{ey}}   &  0  \\
 0  &  E^{\epsilon}_{h}-\frac{\hbar^{2}k^{2}_{x}}{2m^{\epsilon}_{hx}}-\frac{\hbar^{2}k^{2}_{y}}{2m^{\epsilon}_{hy}}   \\
\end{array}
\right),\label{eq.14}
\end{eqnarray}
where $E^{\epsilon}_{e}=u^{\epsilon}+\delta^{\epsilon}$,
$E^{\epsilon}_{h}=u^{\epsilon}-\delta^{\epsilon}$ and
\begin{eqnarray}
  m^{\epsilon}_{ex}&=&\frac{\hbar^{2}}{2(\eta^{\epsilon}_{x}+\gamma^{\epsilon}_{x}
 +(\chi^{\epsilon})^{2}/2\delta^{\epsilon})} \nonumber \\
 m^{\epsilon}_{ey}&=&\frac{\hbar^{2}}{2(\eta^{\epsilon}_{y}+\gamma^{\epsilon}_{y})} \nonumber \\
 m^{\epsilon}_{hx}&=&\frac{\hbar^{2}}{2(\gamma^{\epsilon}_{x}-\eta^{\epsilon}_{x}
 +(\chi^{\epsilon})^{2}/2\delta^{\epsilon}))}  \nonumber \\
 m^{\epsilon}_{hy}&=&\frac{\hbar^{2}}{2(\gamma^{\epsilon}_{y}-\eta^{\epsilon}_{y})} .\label{eq.15}
\end{eqnarray}

In the reminder of this section we consider the effects of the
strains applied along all three principle directions of
phosphorene on its electronic band structure, as a key feature of
crystalline materials to explore their other physical properties.
We have two aims. One is to see whether our tight-binding
Hamiltonian reproduces previous
results\cite{Rodin1,Han1,Qin2,Peng1,Jiang4} for the energy gap of
strained phosphorene. The other is to show that in what energy
region the decoupled energy bands agree well with the others,
obtained from the low-energy and the tight-binding Hamiltonian.

%%%%%%%%%%%%%%%%%%%%%%%%%%%%%%%%%%%%%%%%%%%%%%%%%%%%%%%%%%%%%%%%%
\textit{Uniaxial strain along the normal direction (z-axis)-} Let
us first explore effects of a uniaxial strain in the normal
direction (z-direction), $\epsilon_{x}=\epsilon_{y}=0$ and
$\epsilon_{z}\neq0$. If we recall the relations obtained for the
the strained bond lengths and the transfer energies, and
substitute them into Eq.~\ref{eq.12}, we get
\begin{eqnarray}
\Delta
E_{g}&=&-(8t_{1}\frac{z^{2}_{1}}{r^{2}_{1}}+4t_{2}\frac{z^{2}_{2}}{r^{2}_{2}}+
8t_{3}\frac{z^{2}_{3}}{r^{2}_{3}}+4t_{5}\frac{z^{2}_{5}}{r^{2}_{5}})\epsilon_{z}
=-12.693\epsilon_{z},\label{eq.16}
\end{eqnarray}
for the strain-induced modulation in the energy gap. This shows
that the energy gap of phosphorenre decreases (increases) linearly
when it is exposed to a uniaxial tensile (compressive) strain in
the normal direction. This is in agreement with the previous
first-principle\cite{Rodin1,Han1,Qin2} and tight-binding
\cite{Jiang4} studies on the strain-induced modulation in the
energy gap of strained phosphorene, done in the linear deformation
regime. This can also be seen in Fig. \ref{fig.03} where we have
shown the energy bands of strained phosphorene for different
values of $\epsilon_{z}$ obtained by diagonalizing Eq. \ref{eq.11}
(black curves), Eq. \ref{eq.13} (red dashed curves) and Eq.
\ref{eq.14} (green dotted-dashed curves). In this figure right
(left) panels show the energy bands of phosphorene in the presence
of a uniaxial tensile (compressive) strain applied in the normal
direction, and in each panel the energy bands have been drown in
both $\Gamma-X$ and $\Gamma-Y$ directions.

This figure also shows that a uniaxial tensile (compressive)
strain in the normal direction enhances (weakens) the anisotropy
of the both electron and hole energy bands slightly (See black
cures in Fig. \ref{fig.03}). This becomes more clear in the next
section, where we consider the strain dependence of the carrier
mobility in strained phosphorene. Moreover one can see that in the
presence of a uniaxial tensile (compressive) strain, the energy
range in which the decoupled bands agree well with the other bands
becomes limited (extended). This is mainly due to the effect of
the stain on the energy gap (See Eq. \ref{eq.16}). When the energy
gap increases (decreases), the coupling of the conduction and the
valance band is enhanced (weakened) and the decoupling-band
approximation becomes more (less) accurate. For a uniaxial tensile
(compressive) strain about $\epsilon_{z}=0.04$
($\epsilon_{z}=-0.04$), the decoupled conduction band overlap well
with the tight-banding conduction band up to $0.11~eV$ ($0.18~eV$)
with respect to the bottom of the conduction band. By making use
of $n=\frac{m^{\epsilon}_{eff}}{\pi \hbar^{2}}E^{\epsilon}_{F}$,
where $E^{\epsilon}_{F}$ is counted from the bottom (top) of the
conduction (valance) band, one can show this agreement corresponds
to $n=1.30\times10^{13}~cm^{-2}$ ($n=3.25\times10^{13}~cm^{-2}$)
electron density. This agreement for the valance band is up to
$0.10~eV$ ($0.16~eV$) with respect to the top of the valance band,
corresponding to $n=1.48\times10^{13}~cm^{-2}$
($n=3.60\times10^{13}~cm^{-2}$) hole density.

%%%%%%%%%%%%%%%%%%%%%%%%%%%%%%%%%%%%%%%%%%%%%%%%%%%%%%%%%%%%%
\textit{Uniaxial strain along the zigzag edge (y-axis)-} When
phosphorene is exposed to a uniaxial strain along its zigzag edge,
the strain-induced modulation in its energy gap is given by
\begin{eqnarray}
\Delta
E_{g}&=&-(8t_{1}\frac{y^{2}_{1}}{r^{2}_{1}}+4t_{2}\frac{y^{2}_{2}}{r^{2}_{2}}+
8t_{3}\frac{y^{2}_{3}}{r^{2}_{3}}+4t_{5}\frac{y^{2}_{5}}{r^{2}_{5}})\epsilon_{y}
=5.945\epsilon_{y},\label{eq.17}
\end{eqnarray}
which shows that a uniaxial tensile (compressive) strain in the
zigzag edge increases (decreases) linearly the energy gap. This
agrees well with the recent studies\cite{Liu1,Peng1,Jiang4}.
Figure \ref{fig.04} shows that in the presence of a uniaxial
tensile (compressive) strain along the zigzag edge, the anisotropy
of the band structure is weakened (enhanced). Moreover it is
evident that, in the presence of a uniaxial tensile (compressive)
strain about $\epsilon_{y}=0.04$ ($\epsilon_{y}=-0.04$), there is
good agreement between the decoupled and the tight-binding
conduction bands up to $0.16~eV$ ($0.13~eV$) with respect to the
bottom of the conduction band. In the valance band the overlapping
is up to $0.15~eV$ ($0.12~eV$) with respect to the top of the
valance band.

%%%%%%%%%%%%%%%%%%%%%%%%%%%%%%%%%%%%%%%%%%%%%%%%%%%%%%%%%%%%%%%%%%
\textit{Uniaxial strain along the armchair edge (x-axis)-} In the
presence of a uniaxial along the armchair edge of phosphorene, the
strain-induced modulation in its energy gap is given by
\begin{eqnarray}
\Delta
E_{g}&=&-(8t_{1}\frac{x^{2}_{1}}{r^{2}_{1}}+4t_{2}\frac{x^{2}_{2}}{r^{2}_{2}}+
8t_{3}\frac{x^{2}_{3}}{r^{2}_{3}}+4t_{5}\frac{x^{2}_{5}}{r^{2}_{5}})\epsilon_{x}
=3.708\epsilon_{x}.\label{eq.18}
\end{eqnarray}
which shows that the energy gap is a linear function of the
applied strain, and increases (decreases) when phosphorene is
exposed to a uniaxial tensile (compressive) strain in agreement
with the recent studies\cite{Liu1,Jiang4}. Comparison of Eqs.
\ref{eq.17} and \ref{eq.18} shows that, for same uniaxial strains
along the zigzag and armchair edges, the uniaxial strain along the
zigzag edge induces a larger band gap variation. Effects of the
applied strain on the anisotropy of the band structure can be seen
in Fig. \ref{fig.05} which, as it is expected, is unlike the
effects of the uniaxial strain along the armchair edges. In the
presence of a uniaxial tensile strain about $\epsilon_{y}=0.04$
along the zigzag edge, the overlapping of the decoupled band with
the tight-binding is up $0.16~eV$ and $0.14~eV$ for the conduction
and the valance bands respectively, while for a compressive strain
about $\epsilon_{y}=-0.04$ they agree only up $0.13~eV$ for both
conduction and valance bands.

We end this section by this conclusion that the electronic band
structure of strained phosphorene, for the experimentally
accessible carrier densities and the uniaxial strains applied
along all three principle directions of phosphorene, is well
described by the decoupled Hamiltonian. Motivated by this fact, we
apply it to consider strain engineering the
charged-impurity-limited carrier mobility in phosphorene.

\section{Strain engineering the charged-impurity-limited carrier mobility}
\label{sec:4}

In this section, employing our strain-dependent decoupled
Hamiltonian, we investigate the strain dependence of the
impurity-limited carrier mobility in phosphorene for both types of
carriers, electron and hole, and along both armchair and zigzag
edges. The carrier mobility, $\mu$, is defined as $\mu=\sigma/ne$
where $\sigma$ is the electrical conductivity, $n$ is the carrier
density and $e$ is the electron charge. To calculate the
electrical conductivity we use the semi-classical Boltzmann
transport theory combined with the relaxation time approximation.
Moreover we restrict our calculation to the steady state and
suppose that the two-dimensional electron gas in phosphorene is
homogenous, so the electrical conductivity is given by
\begin{eqnarray}
\sigma_{ii}=-e^{2}g_{s}\int\frac{d^{2}\mathbf{k}}{(2\pi)^{2}}
\tau(E_{\mathbf{k}})v_{i}(\mathbf{k})\frac{\partial
f(E_{\mathbf{k}})}{\partial E_{\mathbf{k}}},\label{eq.19}
\end{eqnarray}
where $i$ is $x,y$, $g_{s}=2$ is the spin degeneracy,
$\mathbf{k}=(k_{x},k_{y})$ is the two-dimensional momentum and
$v_{i}=\hbar k_{i}/m^{\epsilon}_{i}$ is the electron velocity in
the $i$ direction with $k_{i}$ and $m^{\epsilon}_{i}$ being the
corresponding electron or hole momentum and mass. $E_{\mathbf{k}}$
is the energy band obtained from the strained-dependent decoupled
Hamiltonian (Notice we have omitted the electron and hole indexes
in $m^{\epsilon}_{i}$ and $E_{\mathbf{k}}$.), $f(E_{\mathbf{k}})$
is the Fermi-Dirac distribution function and
$\tau(E_{\mathbf{k}})$ is the relaxation time. Let us suppose that
the impurities are static, of symmetric potential and have no
internal excitations. So the relaxation time is given by
\begin{eqnarray}
\frac{1}{\tau(E_{\mathbf{k}})}=\frac{2\pi
n_{i}}{\hbar}\int\frac{d^{2}\mathbf{k}^{'}}{(2\pi)^{2}}
|\frac{V_{i}(q)}{\varepsilon(q)}|^{2}(1-\cos\theta_{\mathbf{k}
\mathbf{k}^{'}})\delta(E_{k}-E_{k^{'}}),\label{eq.20}
\end{eqnarray}
where $n_{i}$ is the number of impurities per unit area,
$q=|\mathbf{k}-\mathbf{k}^{'}|$ and $\theta_{\mathbf{k}
\mathbf{k}^{'}}$ is the scattering angle between $\mathbf{k}$ and
$\mathbf{k}^{'}$. $V_{i}(q)=\frac{2\pi e^{2}}{\kappa q}$ is the
Fourier transform of the potential of the charge impurity and
$\kappa=(\kappa_{sub}+\kappa_{enc})/2$ is the effective dielectric
constant with $\kappa_{sub}$ and $\kappa_{enc}$ being the
dielectric constant of the substrate ($\kappa_{sub}=2.5$ for
$SiO_{2}$ substrate\cite{Low1}) and the encapsulating layer
respectively which for vacuum is zero. $\varepsilon(q)$ is the
dielectric function which within the random phase approximation is
given by $\varepsilon(q)=1+\frac{2\pi e^{2}}{\kappa q}\Pi(q)$,
where $\Pi(q)$ is the polarizability function. The polarizability
function can be written\cite{Low1} as
\begin{eqnarray}
\Pi(q)=\frac{\sqrt{m^{\epsilon}_{x}m^{\epsilon}_{y}}}{\pi
\hbar^{2}}
\left[1-\sqrt{1-\frac{8E^{\epsilon}_{F}/\hbar^{2}}{q_{x}^{2}/m^{\epsilon}_{x}
+q_{y}^{2}/m^{\epsilon}_{y}}}\right],\label{eq.21}
\end{eqnarray}
where $E^{\epsilon}_{F}$ is the Fermi energy of strained
phosphorene for a fixed carrier concentration.

If we introduce new variables as
$p_{x}=(\frac{m^{\epsilon}_{y}}{m^{\epsilon}_{x}})^{1/4}k_{x}$ and
$p_{y}=(\frac{m^{\epsilon}_{x}}{m^{\epsilon}_{y}})^{1/4}k_{y}$, we
have $E_{p}=\frac{1}{2m^{\epsilon}_{eff}}(p_{x}^{2}+p_{y}^{2})$
for the energy bands with
$m^{\epsilon}_{eff}=\sqrt{m^{\epsilon}_{x}m^{\epsilon}_{y}}$. In
the new momentum, space the electrical conductivity is given by
\begin{eqnarray}
\sigma_{ii}=\frac{m_{eff}}{m^{\epsilon}_{i}}
\frac{e^{2}g_{s}}{2\pi \hbar^{2}} \int E_{p} dE_{p} \tau(E_{p})
(-\frac{
\partial f(E_{p}) }{\partial E_{p}}),\label{eq.22}
\end{eqnarray}
leading to
$\sigma_{ii}=\frac{g_{s}m^{\epsilon}_{eff}}{m^{\epsilon}_{i}}
\frac{e^{2}}{h}\frac{E^{\epsilon}_{F}\tau(E^{\epsilon}_{F})}{\hbar}$
for the electrical conductivity of strained phosphorene at zero
temperature, where
$E^{\epsilon}_{F}=\frac{\hbar^{2}(p^{\epsilon}_{F})^{2}}{2m^{\epsilon}_{eff}}$
with $p^{\epsilon}_{F}$ being Fermi momentum in the new momentum
space. $\tau(E^{\epsilon}_{F})$ is given by
\begin{eqnarray}
\frac{1}{\tau(E^{\epsilon}_{F})}&=&\frac{n_{i}m^{\epsilon}_{eff}}{\pi\hbar^{3}}
\int_{0}^{\pi} d\theta \left| \frac{2\pi e^{2}}{\sqrt{2}\kappa
k_{F}\sqrt{1-\cos\theta}+2\pi e^{2}D(E^{\epsilon}_{F})}
\right|^{2}(1-\cos\theta),\label{eq.23}
\end{eqnarray}
where
$D(E^{\epsilon}_{F})=\frac{m^{\epsilon}_{eff}}{\pi\hbar^{2}}$ is
the carrier density of states at the Fermi energy and
\begin{eqnarray}
k_{F}=\sqrt{2\pi n \left [
(\frac{m^{\epsilon}_{y}}{m^{\epsilon}_{x}})^{1/2}\cos^{2}\theta
+(\frac{m^{\epsilon}_{x}}{m^{\epsilon}_{y}})^{1/2}\sin^{2}\theta
\right ]},\label{eq.24}
\end{eqnarray}
is the anisotropic Fermi momentum. In Eq. \ref{eq.24}, $\theta$ is
counted from the x-axis, and $n=\frac{m^{\epsilon}_{eff}}{\pi
\hbar^{2}}E^{\epsilon}_{F}$ is the carrier density in strained
phosphorene, being a linear function of Fermi energy as same as
the carrier density in the ordinary two-dimensional electron gas.
Hence the zero-temperature carrier mobility in strained
phosphorene is given by$\mu_{ii}=\frac{e
\tau(E^{\epsilon}_{F})}{m^{\epsilon}_{i}}$.

In Fig. \ref{fig.06} we have shown our numerical results for the
strain dependence of the charged-impurity-limited electron (left
panels) and hole (right panels) in phosphorene exposed to the
uniaxial strains in the normal direction (z-axis). The upper
(lower) panels shows the carrier mobility along its armchair
(zigzag) edge, and orange ($n=0.2\times10^{13}cm^{-2}$) to black
($n=1.0\times10^{13}cm^{-2}$) curves correspond to different
carrier densities with $\Delta n=0.2\times10^{13}cm^{-2}$. The
density of the charged impurities is supposed to be
$n_{i}=1.0\times 10^{13} cm^{-1}$ which is typical of the
$SiO_{2}$ substrate. Figure \ref{fig.06} shows that the carrier
mobility along the armchair direction is higher than that along
the zigzag direction, as same as that in unstrained
phosphorene\cite{Liu1,Ong1,Gillgren1,Fei2}. This can be understood
by this fact that, in the presence of both uniaxial tensile and
compressive strains, the carrier effective mass along the armchair
edge is always smaller than that along the zigzag edge. This can
be tested by making use of the Eqs. \ref{eq.15}, \ref{eq.A1}. and
\ref{eq.A2}. Moreover one can see that the carrier mobility along
both armchair and zigzag directions increases by increasing the
carrier density. This is the familiar feature of the ordinary
two-dimensional electron gas\cite{Ando1}, arising from the linear
dependence of its carrier density on the Fermi energy (In
phosphorene the carrier density depends on the Fermi energy as
$n=\frac{m^{\epsilon}_{eff}}{\pi \hbar^{2}}E^{\epsilon}_{F}$).
Figure \ref{fig.06} also shows that in the presence of a tensile
(compressive) strain in the normal direction, the carrier mobility
along the armchair edge increases (decreases), while the carrier
mobility along the zigzag edge decreases (increases). This
property originates from the effect of the strain on the
anisotropy (and consequently the carrier effective mass) in
phosphorene, as explained in the previous section. To explain this
property further, we rewrite the relation of the carrier mobility
as $\mu_{ii}=e \frac{m^{\epsilon}_{eff}}{m^{\epsilon}_{i}}
\frac{\tau(E^{\epsilon}_{F})}{m^{\epsilon}_{eff}}$. It is easy to
show that the effect of the strain on
$\frac{\tau(E^{\epsilon}_{F})}{m^{\epsilon}_{eff}}$ part is weak
and it mainly affects on
$\frac{m^{\epsilon}_{eff}}{m^{\epsilon}_{i}}$ part. By making use
of the Eqs. \ref{eq.15}, \ref{eq.A1}. and \ref{eq.A2}, one can
show that applying a uniaxial tensile (compressive) strain in the
normal direction decreases (increases) both electron and hole
effective masses in the armchair (zigzag) direction, and
consequently their mobilities in the armchair (zigzag) direction
increase (decrease).

In Fig. \ref{fig.07} we have compared the effect of the direction
of the applied straina on the carrier mobility in phosphorene.
Figure \ref{fig.07} shows that, unlike the strains in the normal
direction, applying a uniaxial tensile (compressive) strain in the
zigzag direction decreases (increases) both electron and hole
mobilities in the armchair (zigzag) direction. This originates
from their different effects on the anisotropy (and consequently
the carrier effective mass) in phosphorene, as explained in the
previous section and above. This figure also shows that applying a
uniaxial strain in the armchair direction weakly change the
carrier mobility in phosphorene. Moreover Figs. \ref{fig.06} and
\ref{fig.07} show that applying a uniaxial tensile (compressive)
strain in the normal (zigzag) direction enhances (weakens) the
anisotropy of the carrier mobility in phosphorene. While in the
presence of a uniaxial compressive (tensile) strain in the normal
(zigzag) direction, the carrier mobility is weakened (enhanced).

\section{Summary and conclusions}
\label{sec:5}

In Summary, we investigated the electronic band structure of
strained phosphorene within the linear deformation regime and
based on the tight-binding model. We restricted our consideration
to the uniaxial strains applied along one of the principle
directions of phosphorene, the normal, the armchair and the zigzag
directions. We showed that the derived strain-dependent energy
spectrums reproduce the previous results for the energy gap of
strained phosphorene. Then we applied the continuum approximation
to derive the corresponding low-energy Hamiltonian. Moreover we
showed, when the interband coupling is weak, the low-energy
Hamiltonian can project into a decoupled electron-hole
Hamiltonian. We found that the electronic band structure of the
strained phosphorene, for the experimentally accessible carrier
densities and the mechanical strains, is well described by the
decoupled Hamiltonian. Motivated by this fact we used our
strain-dependent decoupled Hamiltonian to investigate the strain
dependence of the charged-impurity-limited carrier mobility in
phosphorene. We examined the dependence of carrier mobility on the
direction of mobility, the carrier type, the carrier density and
the direction of the applied strain. We showed the dependence of
the carrier mobility on the direction of mobility, the carrier
type and the carrier density is same as that in unstrained
phosphorene. Moreover, as a point worthy of mention, we found that
applying a uniaxial tensile (compressive) strain in the normal
direction decreases (increases) carrier mobility in the armchair
(zigzag) direction. While in the presence of a uniaxial tensile
(compressive) strain in the zigzag direction the carrier mobility
is decreased (increased). We also showed that a uniaxial strain in
the armchair direction don't changed the carrier mobility
approximately. These properties were explained based on the effect
of the applied strain on the anisotropy of the carrier effective
mass in phosphorene.

%%%%%%%%%%%%%%%%%%%%%%%%%%%%%%%%%%%%%%%%%%%%%%%%%%%%%%%%%%%%%%%%%%
%%%%%%%%%%%%%%%%%%%%%%%%%%%%%%%%%%%%%%%%%%%%%%%%%%%%%%%%%%%%%%%%%%
\appendix
\section{Appendix: Calculating the elements of the strain-dependent low-energy Hamiltonian matrix}
The matrix element in Eq. \ref{eq.13}, in the linear deformation
regime, depend in general on the applied strain as
\begin{eqnarray}
u^{\epsilon}=u+\epsilon u^{'} \nonumber \\
\eta^{\epsilon}_{x}=\eta_{x}+\epsilon \eta^{'}_{x} \nonumber \\
\eta^{\epsilon}_{y}=\eta_{y}+\epsilon \eta^{'}_{y} \nonumber \\
\delta^{\epsilon}=\delta+\epsilon \delta^{'} \nonumber \\
\gamma^{\epsilon}_{x}=\gamma_{x}+\epsilon \gamma^{'}_{x}  \nonumber \\
\gamma^{\epsilon}_{y}=\gamma_{y}+\epsilon \gamma^{'}_{y}  \nonumber \\
 \chi^{\epsilon}=\chi+\epsilon \chi^{'},\label{eq.A1}  \end{eqnarray}
where the coefficients of the applied strain, for a uniaxial
strain in the normal direction, are given by
\begin{eqnarray}
u^{'}&=&-8t_{4}\frac{z_{4}^{2}}{r_{4}^{2}}=0.32~eV \nonumber \\
\eta^{'}_{x}&=&4t_{4}\frac{z_{4}^{2}x_{4}^{2}}{r_{4}^{2}}=-0.75~eV\AA^{2} \nonumber \\
\eta^{'}_{y}&=&4t_{4}\frac{z_{4}^{2}y_{4}^{2}}{r_{4}^{2}}=-0.43~eV\AA^{2} \nonumber \\
\delta^{'}&=&-4t_{1}\frac{z^{2}_{1}}{r^{2}_{1}}-2t_{2}\frac{z^{2}_{2}}{r^{2}_{2}}
-4t_{3}\frac{z^{2}_{3}}{r^{2}_{3}}-2t_{5}\frac{z^{2}_{5}}{r^{2}_{5}}=-6.58~eV \nonumber \\
\gamma^{'}_{x}&=&2t_{1}\frac{z^{2}_{1}x^{2}_{1}}{r^{2}_{1}}+t_{2}\frac{z^{2}_{2}x^{2}_{2}}{r^{2}_{2}}
+2t_{3}\frac{z^{2}_{3}x^{2}_{3}}{r^{2}_{3}}+t_{5}\frac{z^{2}_{5}x^{2}_{5}}{r^{2}_{5}}=1.44~eV\AA^{2} \nonumber \\
\gamma^{'}_{y}&=&2t_{1}\frac{z^{2}_{1}y^{2}_{1}}{r^{2}_{1}}+2t_{3}\frac{z^{2}_{3}y^{2}_{3}}{r^{2}_{3}}=0.00~eV\AA^{2} \nonumber \\
\chi^{'}&=&-4t_{1}\frac{z^{2}_{1}x_{1}}{r^{2}_{1}}-2t_{2}\frac{z^{2}_{2}x_{2}}{r^{2}_{2}}
-4t_{3}\frac{z^{2}_{3}x_{3}}{r^{2}_{3}}-2t_{5}\frac{z^{2}_{5}x_{5}}{r^{2}_{5}}
=4.74~eV\AA.\label{eq.A2}
\end{eqnarray}
These coefficients, when the strain is applied in the zigzag edge
(y-axis), become
\begin{eqnarray}
u^{'}&=&-8t_{4}\frac{y_{4}^{2}}{r_{4}^{2}}=0.19~eV \nonumber \\
\eta^{'}_{x}&=&4t_{4}\frac{y_{4}^{2}x_{4}^{2}}{r_{4}^{2}}=-0.45~eV\AA^{2} \nonumber \\
\eta^{'}_{y}&=&4t_{4}\frac{y_{4}^{4}}{r_{4}^{2}}-4t_{4}y_{4}^{2}=0.89~eV\AA^{2} \nonumber \\
\delta^{'}&=&-4t_{1}\frac{y^{2}_{1}}{r^{2}_{1}}-2t_{2}\frac{y^{2}_{2}}{r^{2}_{2}}
-4t_{3}\frac{y^{2}_{3}}{r^{2}_{3}}-2t_{5}\frac{y^{2}_{5}}{r^{2}_{5}}=2.91~eV \nonumber \\
\gamma^{'}_{x}&=&2t_{1}\frac{y^{2}_{1}x^{2}_{1}}{r^{2}_{1}}+t_{2}\frac{y^{2}_{2}x^{2}_{2}}{r^{2}_{2}}
+2t_{3}\frac{y^{2}_{3}x^{2}_{3}}{r^{2}_{3}}+t_{5}\frac{y^{2}_{5}x^{2}_{5}}{r^{2}_{5}}=-3.81~eV\AA^{2} \nonumber \\
\gamma^{'}_{y}&=&2t_{1}\frac{y^{4}_{1}}{r^{2}_{1}}-2t_{1}y^{2}_{1}
+2t_{3}\frac{y^{4}_{3}}{r^{2}_{3}}-2t_{3}y^{2}_{3}=3.81~eV\AA^{2} \nonumber \\
\chi^{'}&=&-4t_{1}\frac{y^{2}_{1}x_{1}}{r^{2}_{1}}-2t_{2}\frac{y^{2}_{2}x_{2}}{r^{2}_{2}}
-4t_{3}\frac{y^{2}_{3}x_{3}}{r^{2}_{3}}-2t_{5}\frac{y^{2}_{5}x_{5}}{r^{2}_{5}}
 =3.43~eV\AA,\label{eq.A3}
\end{eqnarray}
while for a uniaxial strain along the armchair edge, they are
\begin{eqnarray}
u^{'}&=&-8t_{4}\frac{x_{4}^{2}}{r_{4}^{2}}=0.33~eV \nonumber \\
\eta^{'}_{x}&=&4t_{4}\frac{x_{4}^{4}}{r_{4}^{2}}-4t_{4}x_{4}^{2}=1.21~eV\AA^{2} \nonumber \\
\eta^{'}_{y}&=&4t_{4}\frac{x_{4}^{2}y_{4}^{2}}{r_{4}^{2}}=-0.45~eV\AA^{2} \nonumber \\
\delta^{'}&=&-4t_{1}\frac{x^{2}_{1}}{r^{2}_{1}}-2t_{2}\frac{x^{2}_{2}}{r^{2}_{2}}
-4t_{3}\frac{x^{2}_{3}}{r^{2}_{3}}-2t_{5}\frac{x^{2}_{5}}{r^{2}_{5}}=2.15~eV \nonumber \\
\gamma^{'}_{x}&=&2t_{1}\frac{x^{4}_{1}}{r^{2}_{1}}-2t_{1}x_{1}^{2}
+t_{2}\frac{x^{4}_{2}}{r^{2}_{2}}-t_{2}x_{2}^{2}
+2t_{3}\frac{y^{2}_{3}x^{2}_{3}}{r^{2}_{3}}-2t_{3}x_{3}^{2}
+t_{5}\frac{y^{2}_{5}x^{2}_{5}}{r^{2}_{5}}-t_{5}x_{5}^{2}
=2.37~eV\AA^{2} \nonumber \\
\gamma^{'}_{y}&=&2t_{1}\frac{x^{2}_{1}y^{2}_{1}}{r^{2}_{1}}+2t_{3}\frac{x^{2}_{3}y^{2}_{3}}{r^{2}_{3}}=-3.81~eV\AA^{2} \nonumber \\
\chi^{'}&=&-4t_{1}\frac{x^{3}_{1}}{r^{2}_{1}}+2t_{1}x_{1}
-2t_{2}\frac{x^{3}_{2}}{r^{2}_{2}}+t_{2}x_{2}
-4t_{3}\frac{x^{3}_{3}}{r^{2}_{3}}+2t_{3}x_{3}
-2t_{5}\frac{x^{3}_{5}}{r^{2}_{5}}+t_{5}x_{5}
=-2.96~eV\AA.\label{eq.A4}
\end{eqnarray}
The lattice parameters, which we used here, are $d_{1}=2.22 \AA$,
$d_{2}=2.24 \AA$, $t_{1}=-1.220 eV$, $t_{2}=3.665 eV$,
$t_{3}=-0.205 eV$, $t_{4 }=-0.105 eV$, $t_{5}=-0.055 eV$
\cite{Rudenko1}, $\alpha=0.2675\pi $ and $\theta=0.567\pi$
\cite{Jiang3}.

%\newpage
%

%
%
%
\newpage
\begin{figure}
\begin{center}
\includegraphics[width=12cm,angle=0]{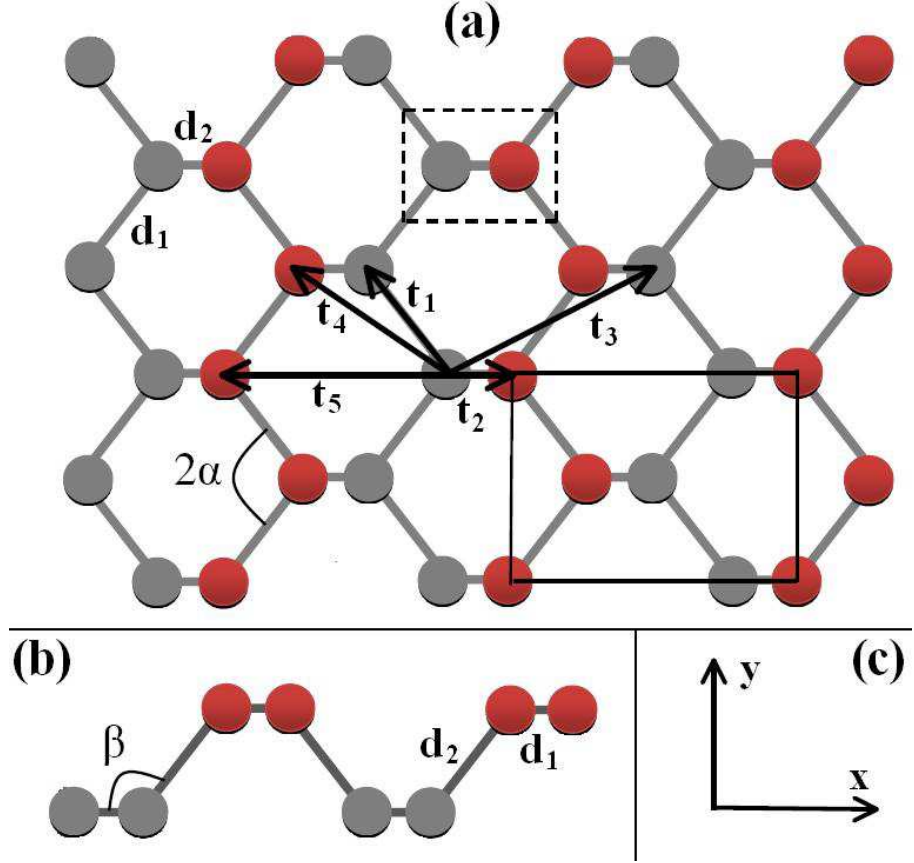}
\caption{(a) The top view of phosphorene lattice structure.
$t_{i}$ indicates to transfer energies from a site to its ith
nearest neighbors. The solid-line (dashed-line) rectangle denotes
to the unit cell in the 4-band (2-band) model. The other
parameters are $d_{1}=2.22 \AA$, $d_{2}=2.24 \AA$,\cite{Rudenko1}
and $\alpha=0.2675 \pi$.\cite{Jiang3} (b) The side view of
phosphorene lattice structure where $\beta=0.567
\pi$.\cite{Jiang3} (c) The coordinate system used in this work.
The armchair edge is supposed to be along the x-axis and the
zigzag edge along the y-axis.}\label{fig.01}
\end{center}
\end{figure}
\begin{figure}
\begin{center}
\includegraphics[width=15cm,angle=0]{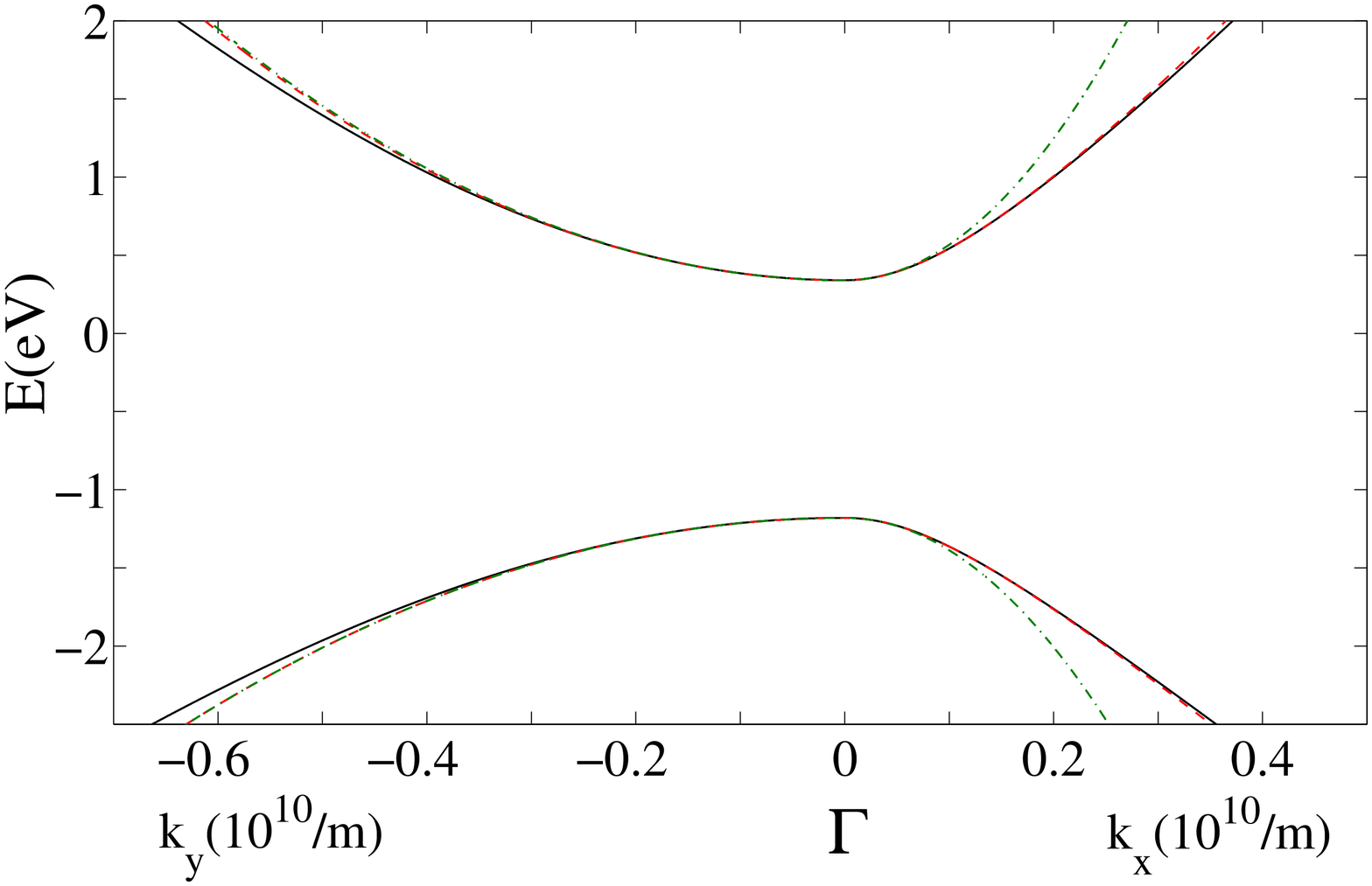}
\caption{The energy spectrums of phosphorenere around $\Gamma$
point obtained from the two-band tight binding Hamiltonian (solid
black cure), the low-energy Hamiltonian (red dashed curve) and the
decoupled Hamiltonian (green dotted-dashed curve).}\label{fig.02}
\end{center}
\end{figure}
\begin{figure}
\begin{center}
\includegraphics[width=15cm,angle=0]{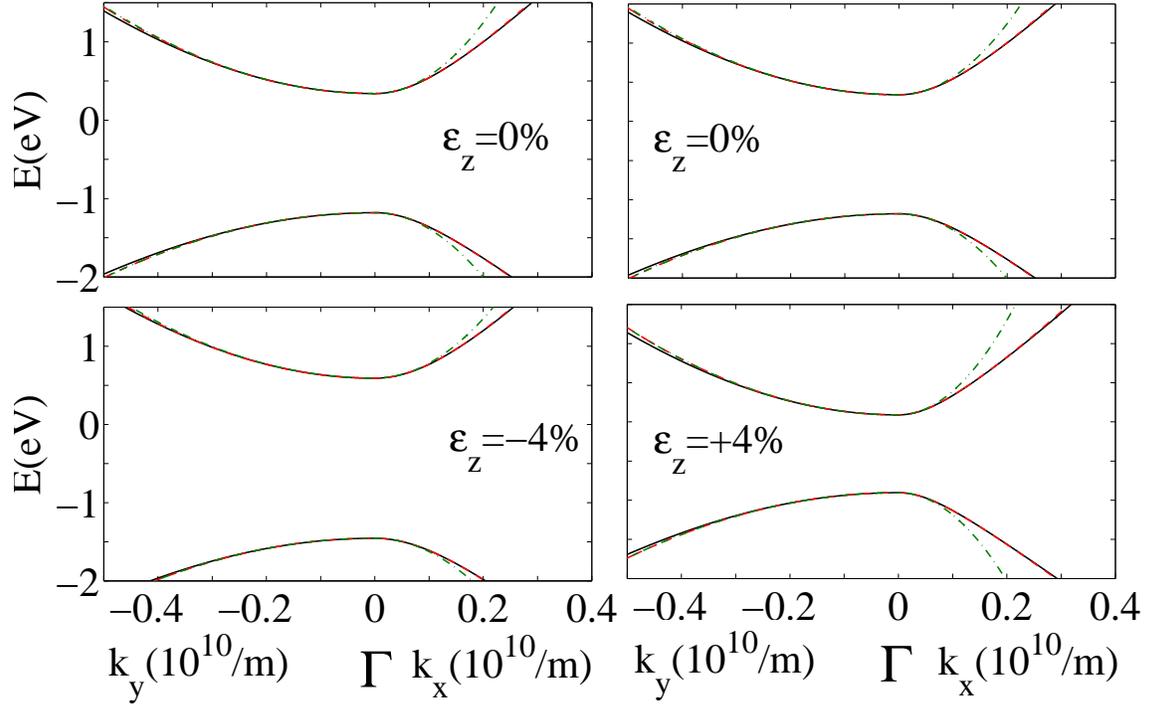}
\caption{The energy spectrums of strained phosphorenere around
$\Gamma$ point obtained from the two-band tight binding
Hamiltonian (black solid cure), the low-energy Hamiltonian (red
dashed curve) and the decoupled electron-hole Hamiltonian (green
dotted-dashed curve). Phosphorene is exposed to uniaxial tensile
(left panels) and compressive (right panels) strains applied in
the normal direction.}\label{fig.03}
\end{center}
\end{figure}
\begin{figure}
\begin{center}
\includegraphics[width=15cm,angle=0]{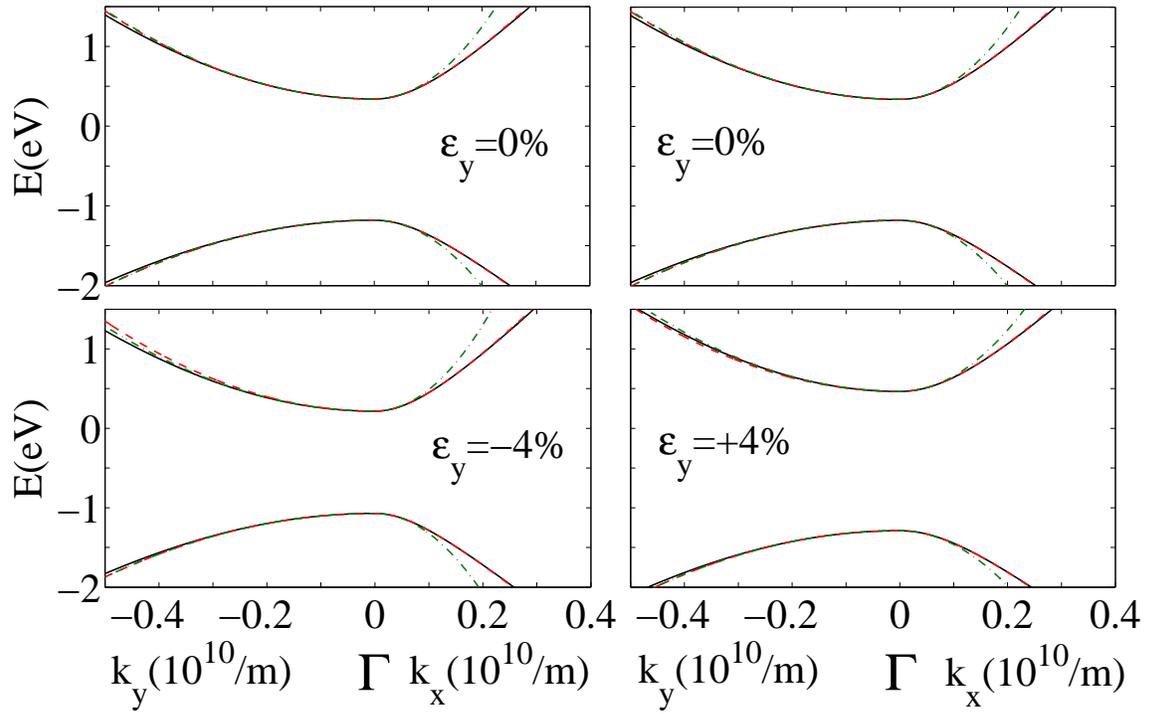}
\caption{Same as Fig. \ref{fig.03} but for strains applied along
the zigzag edge.}\label{fig.04}
\end{center}
\end{figure}
\begin{figure}
\begin{center}
\includegraphics[width=15cm,angle=0]{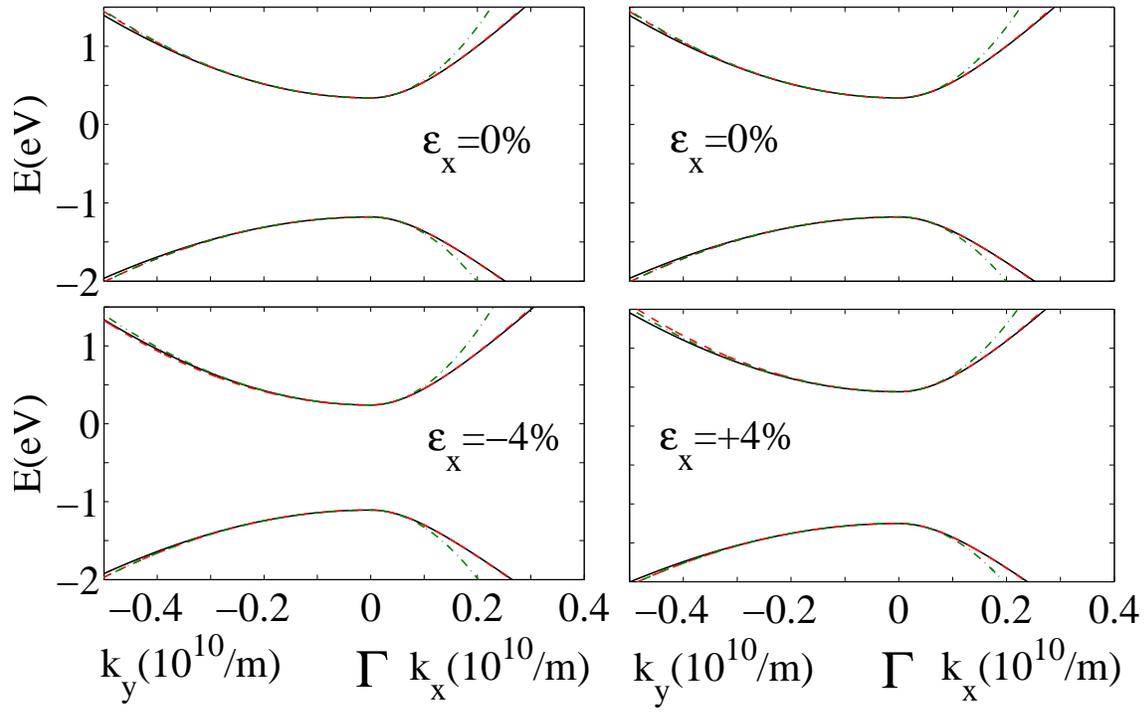}
\caption{Same as Fig. \ref{fig.03} but for strains applied along
the armchair edge.}\label{fig.05}
\end{center}
\end{figure}
\begin{figure}
\begin{center}
\includegraphics[width=15cm,angle=0]{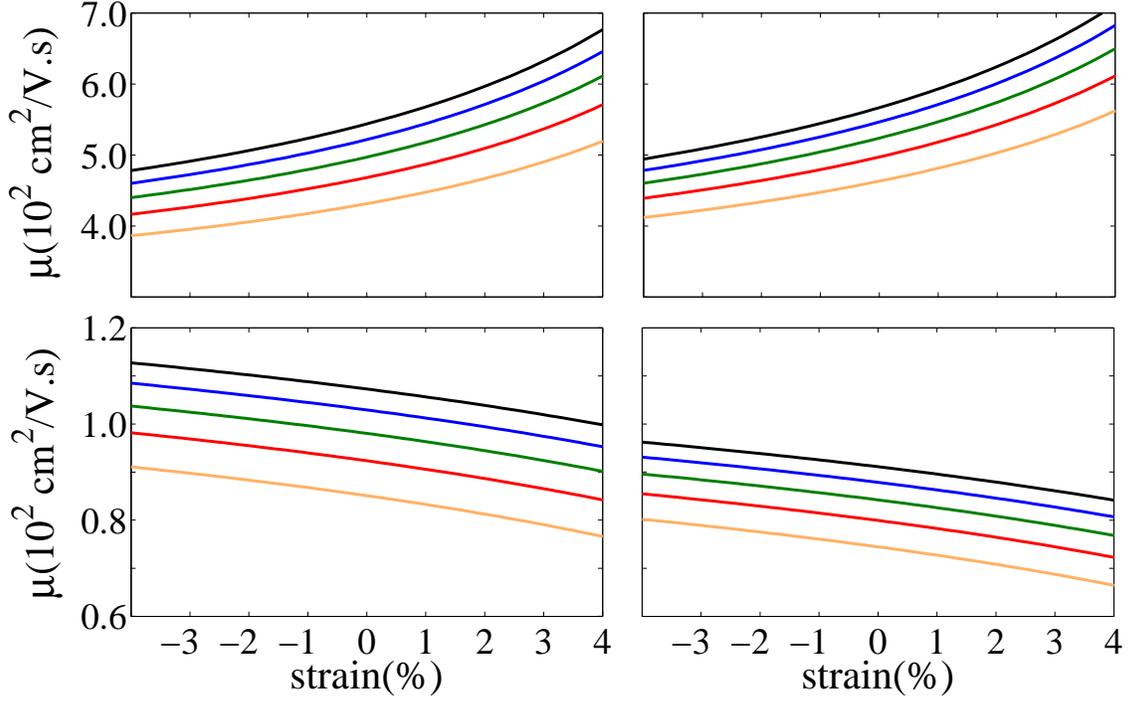}
\caption{The strain dependence of the charged-impurity-limited
electron (left panels) and hole (right panels) mobilities along
the armchair (upper panels) and zigzag (lower panels) edges of
phosphorene exposed to uniaxial strains in the normal direction.
Orange to black lines correspond to $n=0.2\times10^{13}cm^{-2}$ to
$n=1.0\times10^{13}cm^{-2}$ carrier densities with $\Delta
n=0.2\times10^{13}cm^{-2}$ and the density of the charged
impurities is $n_{i}=1.0\times10^{13}cm^{-2}$.}\label{fig.06}
\end{center}
\end{figure}
\begin{figure}
\begin{center}
\includegraphics[width=15cm,angle=0]{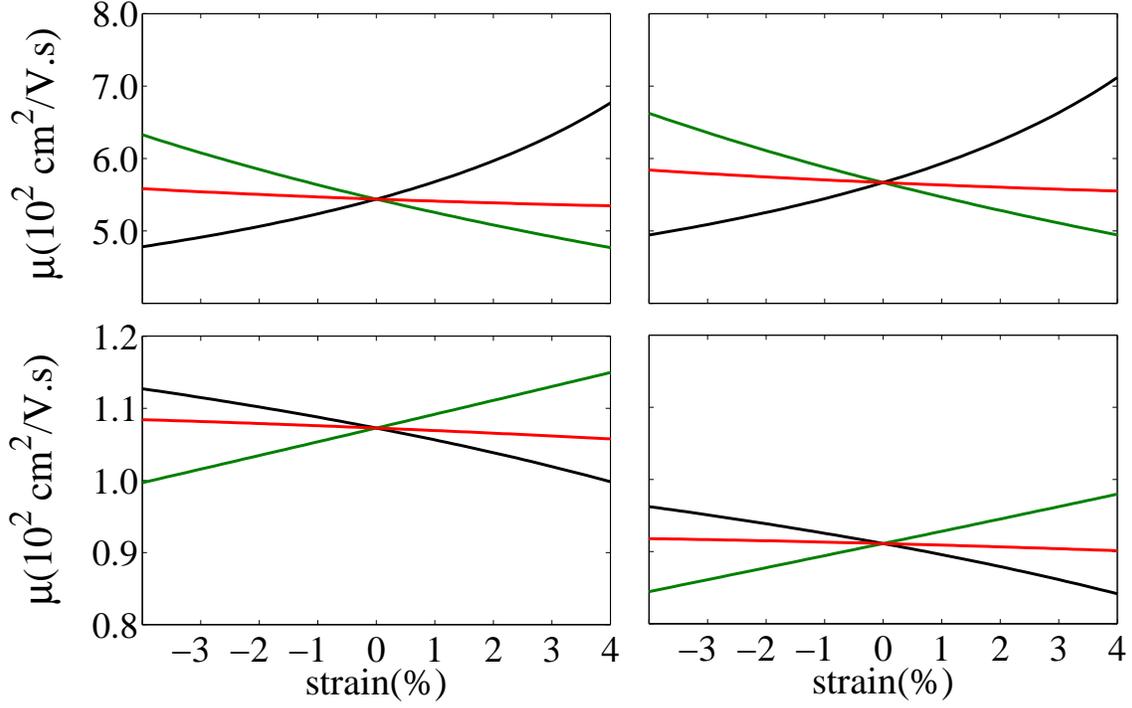}
\caption{The strain dependence of the charged-impurity-limited
electron (left panels) and hole (right panels) mobilities along
the armchair(upper panels) and zigzag(lower panels) edges of
phosphorene, when it is exposed to uniaxial strains along the
normal (black cure), armchair (red cure) and zigzag (green cure)
directions. The electron and hole densities are
$1.0\times10^{13}cm^{-2}$ and the density of the charged
impurities is $n_{i}=1.0\times10^{13}cm^{-2}$.}\label{fig.07}
\end{center}
\end{figure}
\end{document}